\documentclass[twocolumn]{aastex631}

\newcommand{\Fig}{\mbox{Fig.\ }}

\begin{document}

\title{Calcium Excess in Novae: Beyond Nuclear Physics Uncertainties}

\author[0009-0003-6956-076X]{Mallory K. Loria}
\affiliation{Department of Physics \& Astronomy\\
University of Victoria\\
Victoria, BC V8P 5C2, Canada}
\affiliation{TRIUMF\\
4004 Wesbrook Mall \\
Vancouver, BC V6T 2A3, Canada}

\author[0000-0001-6120-3264]{Pavel A. Denissenkov}
\affiliation{Department of Physics \& Astronomy\\
University of Victoria\\
Victoria, BC V8P 5C2, Canada}

\author[0000-0001-6468-6097]{Chris Ruiz}
\affiliation{Department of Physics \& Astronomy\\
University of Victoria\\
Victoria, BC V8P 5C2, Canada}
\affiliation{TRIUMF\\
4004 Wesbrook Mall \\
Vancouver, BC V6T 2A3, Canada}

\author[0000-0001-8087-9278]{Falk Herwig}
\affiliation{Department of Physics \& Astronomy\\
University of Victoria\\
Victoria, BC V8P 5C2, Canada}

\collaboration{10}{(CaNPAN Collaboration, \url{https://canpan.ca}, NuGrid Collaboration, \url{http://nugridstars.org})}



\begin{abstract}

We examine Ca abundances in classical novae from spectroscopic observations spanning 65 years and investigate whether they are systematically high compared to those predicted by nova models. For the first time, we perform Monte Carlo simulations assessing the impact of nuclear reaction rate uncertainties on abundances predicted by multi-zone nova models. While the Ca abundances in the models are sensitive to variations of rates of the reactions \(^{37}\text{Ar}(\textit{p},\gamma)^{38}\text{K}\) and \(^{38}\text{K}(\textit{p},\gamma)^{39}\text{Ca}\), the nuclear physics uncertainties of these reactions cannot account for the discrepancy between the observed and predicted Ca abundances in novae. Furthermore, the overabundance of Ca has important implications for measuring \(^{7}\text{Be}\) in nova ejecta, as Ca lines are used to estimate \(^{7}\text{Be}\) abundances. If the Ca abundance is incorrectly determined, it could lead to inaccurate \(^{7}\text{Be}\) abundance estimates. Possible alternative explanations for the observed Ca overabundance are discussed. 

\end{abstract}



\section{Introduction} \label{sec:intro}

Classical novae are thermonuclear runaways (TNRs) of H burning on accreting white dwarfs (WDs) in close binary systems with a main sequence or evolved companion star \citep[e.g.,][]{prialnik1995,jose1998,jose2008_textbook,truran2008,pavel2014,jose_textbook,jose_2020,starrfield2018}. They are one of the most frequently observed cataclysmic variables, with a few out of a predicted 20--70 eruptions per year observed in the Milky Way galaxy \citep{galaxy_first,galaxy_second}. A vast amount of spectroscopic data exists for novae, but only a few measurements of elemental abundances in the Ca region have been made. Despite the theoretical expectation that novae should not produce Ca, spectral observations exist in which Ca appears to be enhanced compared to its solar value \citep{andrea1994,arkhipova2000,evans2003,morisset1996,Pottasch1959,woodward2021}. The overabundance of Ca in these novae could indicate observational errors, reveal limitations in our nova models, suggest issues with nuclear physics inputs, or point to gaps in our understanding of the nova environment. Moreover, as Ca represents the approximate termination point of nova nucleosynthesis, it serves as a test of our models and understanding of nucleosynthesis in novae. We have tabulated and compared observed abundances of Ca in novae with those from our nova models, along with abundances of other elements measured in the same novae, and conclude there is an excess of Ca. We find a similar situation for Ar. 

The production of intermediate-mass elements in novae was previously explored by \cite{Jose2001}. The authors reported some enhancements of heavier species, such as Ar, K, and, to some extent, Ca were obtained for an extreme nova model with a high peak temperature. The synthesis of Ca was not described in depth. In 2002, sensitivity studies and in 2003 single-zone Monte Carlo (MC) simulations for reaction rate uncertainty studies of nova nucleosynthesis were performed, by \cite{iliadis2002} and \cite{hix2003}. In the first work, individual nuclear reaction rates were multiplied and divided by fixed factors for 142 isotopes, and the impact of those changes on the final abundances was calculated using one-zone nova trajectories. In the second work, all reaction rates relevant to nova nucleosynthesis were randomly varied within their estimated uncertainty ranges, and the impact of those variations on predicted abundances was reported. 

The abundance of \(^{7}\text{Be}\) in novae is traditionally determined using the equivalent widths of \(^{7}\text{Be II}\) and Ca II lines with the assumption that the ionization fractions of Be II/Be and Ca II/Ca are equal. \cite{chugai2020} found that the ionization fraction of Be II/Be in the nova V5668 Sgr should be at least a factor of 10 higher than Ca II/Ca. The authors assumed that the Ca abundance in this nova envelope was solar because Ca is not synthesized in novae. Since the measurements of the \(^{7}\text{Be}\) abundance in novae depend on the strength of Ca spectral lines, if Ca abundances are really enhanced in some novae, this may impact estimates of the \(^{7}\text{Be}\) abundance, which have recently been reported to be too high compared to model predictions \citep[][and references therein]{pavel2021_Be}.

Given the aforementioned potential implications for the Ca overabundance, in this paper we focus exclusively on the nuclear physics aspect, specifically investigating whether uncertainties in nuclear reaction rates activated during nova TNRs can account for the observed Ca overabundances. The remaining possibilities are left to be explored in future studies. In Section \ref{sect:obs} we discuss the observations used in this work along with their uncertainties and limitations. Our nova models will be established as suitable and comprehensive for comparison to observations by verification with previously published results in section \ref{sect: models}. Section \ref{sect: comp w obs} directly compares our nova models with observations and clearly shows the overabundance of Ca. The impact of nuclear physics uncertainties on Ca production and the identification of key reactions whose rate uncertainties are correlated with Ca production are discussed in Section \ref{sect:nphys}. Alternative explanations for the observed overabundance of Ca are presented in Section \ref{sect:disc}. Finally, in Section \ref{sect: concl}, we summarize the main findings of this work.

\section{Observations} \label{sect:obs}

Observations of novae with estimated Ca abundances date back to 1959, when \cite{Pottasch1959} determined the temperatures and radii of central stars for six novae and reported abundances of elements heavier than H. Ca abundances in novae were reported again in 1994 when they were determined for 11 novae using UV and optical spectra \citep{andrea1994}. The authors reported abundance uncertainties of factors 2--3, noting that collisionally excited lines are sensitive to the assumed temperature and density inputs in their models. Shortly after, \cite{morisset1996} used photoionization models to report the Ca abundance in the nova GQ Mus. They estimated that Ca was overabundant by a factor of 3 compared to the solar abundance, but also stressed the importance of using accurate atomic data for reliable abundance measurements. Later, \cite{arkhipova2000} reported abundances for the nova V705 Cas using UV spectra, where Ca was again shown to be overabundant compared to solar. However, the uncertainties in the line intensities were approximately 20--30\%, which were then propagated to large uncertainties in abundances though the authors did not specify the exact magnitude of these abundance uncertainties. \cite{evans2003} presented the results of spectroscopic observations of the nova V7223 Cas, in which temperature and abundance ratios were estimated. In that paper, it was proposed that the overabundances of S and Ca observed in that nova might be associated with its more evolved companion star. 

Observed elemental abundances are typically reported in terms of number densities. To compare these with our simulations, which give abundances in mass fractions, we convert the observed number densities to their corresponding mass fractions. Number densities are reported as ratios with respect to another element, like
\begin{equation}
    \left( \frac{{N}_{\text{Ca}}}{{N}_{\text{H}}} \right)
    \text{or} 
        \left(\frac{{N}_{\text{Ca}}}{{N}_{\text{Si}}}\right).
    	\label{eq:1}
\end{equation}
The number density of an element $i$ is related to its mass fraction as
\begin{equation}
   N_{i} \approx \frac{X_{i} \rho N_{\text{A}}}{A_{i}},
	\label{eq:2}
\end{equation}
where $\rho$ is the mass density, $N_{\text{A}}$ is Avogadro's number, and $A_i$ is the atomic mass of the element. The atomic masses used in these calculations are assumed to be the average atomic masses of stable isotopes with their terrestrial relative abundances. Hence, the ratio of mass fractions for two elements can be written as
\begin{equation}
   \frac{X_{\text{Ca}}}{X_{\text{H}}}  = \frac{N_{\text{Ca}}A_{\text{Ca}}}{N_{\text{H}}A_{\text{H}}}.
    \label{eq:3}
\end{equation}

A detailed breakdown of the abundances, in number density and mass fraction from the observations used in this work is provided in Table \ref{table:combined_obs}.

\begin{table*}[htb!]
\centering
\caption{Observed nova abundances. For each nova, the first row lists number density ratios relative to H, and the second row lists the corresponding mass fractions calculated using Equation \ref{eq:3}.} \label{table:combined_obs}
\resizebox{\textwidth}{!}{\begin{tabular}{cccccccccccc} 
\hline 
\\[0.1ex]
Nova & (N$_{\text{He}}$/N$_{\text{H}}$) & (N$_{\text{C}}$/N$_{\text{H}}$) & (N$_{\text{N}}$/N$_{\text{H}}$) & (N$_{\text{O}}$/N$_{\text{H}}$) & (N$_{\text{Ne}}$/N$_{\text{H}}$) & (N$_{\text{Mg}}$/N$_{\text{H}}$) & (N$_{\text{Si}}$/N$_{\text{H}}$) & (N$_{\text{S}}$/N$_{\text{H}}$) & (N$_{\text{Cl}}$/N$_{\text{H}}$) & (N$_{\text{Ar}}$/N$_{\text{H}}$) & (N$_{\text{Ca}}$/N$_{\text{H}}$) \\
\\[0.1ex]
\hline 
\\[0.5ex]
\multicolumn{12}{c}{\cite{Pottasch1959}} \\ 
\\
V603 Aql & 3.23E-1 & 1.67E-4 & - & 2.22E-2 & 1.43E-4 & - & - & - & - & - & 1.00E-5 \\
X$_{\text{i}}$/X$_{\text{H}}$ & 1.28E0 & 1.99E-3 & - & 3.53E-1 & 2.86E-3 & - & - & - & - & - & 3.98E-4 \\
\\
DQ Her & 7.35E-2 & 5.26E-4 & 3.70E-3 & 3.57E-3 & 1.11E-4 & - & - & - & - & - & 3.13E-5 \\
X$_{\text{i}}$/X$_{\text{H}}$ & 2.92E-1 & 6.27E-3 & 5.15E-2 & 5.67E-2 & 2.22E-3 & - & - & - & - & - & 1.24E-3 \\
\\
GK Per & 1.75E-1 & - & - & 4.10E-3 & 6.67E-4 & - & - & 1.23E-4 & - & - & 1.00E-5 \\
X$_{\text{i}}$/X$_{\text{H}}$ & 6.97E-1 & - & - & 6.50E-2 & 1.33E-2 & - & - & 3.93E-3 & - & - & 3.98E-4 \\
\\
RR Pic & 3.23E-1 & - & - & 1.49E-3 & 5.56E-4 & - & - & 1.20E-4 & - & - & 2.50E-5 \\
X$_{\text{i}}$/X$_{\text{H}}$ & 1.28E0 & - & - & 2.37E-2 & 1.11E-2 & - & - & 3.83E-3 & - & - & 9.94E-4 \\
\\
\hline \\ [0.5ex]
\multicolumn{12}{c}{\cite{andrea1994}} \\
\\
V2214 Oph & 1.90E-1 & - & 6.40E-2 & 1.10E-3 & 2.50E-3 & - & - & 9.70E-5 & - & 5.50E-5 & 9.40E-5 \\
X$_{\text{i}}$/X$_{\text{H}}$ & 7.54E-1 & - & 8.89E-1 & 1.75E-2 & 5.00E-2 & - & - & 3.09E-3 & - & 2.18E-3 & 3.74E-3 \\
\\
V977 Sco & 1.90E-1 & - & 5.90E-3 & 3.70E-3 & 2.50E-3 & - & - & - & - & - & 5.30E-5 \\
X$_{\text{i}}$/X$_{\text{H}}$ & 7.54E-1 & - & 8.20E-2 & 5.87E-2 & 5.00E-2 & - & - & - & - & - & 2.10E-3 \\
\\
V443 Sct & 2.30E-1 & - & 7.80E-3 & 9.00E-4 & 1.40E-5 & - & - & 2.40E-5 & - & 1.80E-5 & 1.00E-5 \\
X$_{\text{i}}$/X$_{\text{H}}$ & 9.13E-1 & - & 1.08E-1 & 1.43E-2 & 2.80E-4 & - & - & 7.64E-4 & - & 7.13E-4 & 3.98E-4 \\
\\
\hline \\ [0.5ex]
\multicolumn{12}{c}{\cite{morisset1996}} \\
\\
GQ Mus & 2.65E-1 & 1.80E-3 & 2.40E-2 & 1.60E-2 & 3.10E-4 & 7.40E-5 & 7.40E-5 & 4.50E-5 & 2.00E-6 & 1.10E-5 & 1.30E-5 \\
X$_{\text{i}}$/X$_{\text{H}}$ & 1.05E0 & 2.14E-2 & 3.34E-1 & 2.54E-1 & 6.21E-3 & 1.78E-3 & 2.06E-3 & 1.43E-3 & 7.03E-5 & 4.35E-4 & 5.21E-4 \\
\\
\hline \\[0.5ex] 
\multicolumn{12}{c}{\cite{arkhipova2000}} \\
\\
V705 Cas & 7.94E-2 & - & 2.51E-2 & 6.31E-3 & - & - & - & - & - & 2.00E-5 & 6.31E-7 \\
X$_{\text{i}}$/X$_{\text{H}}$ & 3.15E-1 & - & 3.49E-1 & 1.00E-1 & - & - & - & - & - & 7.90E-4 & 2.51E-5 \\
\\
\hline \\ [0.5ex]

\multicolumn{12}{c}{Solar Abundance \cite{grevesse1993}} \\
\\

X$_{\text{i}}$/X$_{\text{H}}$ & 3.87E-1 & 4.91E-3 & 1.50E-3 & 1.37E-2 & 2.79E-3 & 1.06E-3 & 1.15E-3 & 5.99E-4 & 1.29E-5 & 1.38E-4 & 1.06E-4 \\
\\
\hline
\end{tabular}}
\end{table*}

Comparing theoretical and observationally derived chemical abundances in novae is complex, owing to limitations in both nova models and spectroscopic analyses. Current nova models are predominantly 1D and assume spherical symmetry, with additional uncertainties coming from input physics, such as nuclear reaction rates. Whereas chemical abundances derived from spectroscopic data carry significant uncertainties due to the challenges of interpreting emission line spectra.

\cite{Helton2012} summarize two principal methods commonly used to derive elemental abundances in nova ejecta: nebular analysis and photoionization modeling. The selection of method depends largely on the evolutionary stage of the nova. Nebular analysis is typically used at later times, once the ejecta have expanded and become optically thin. The authors note that abundances derived using this approach should be considered lower limits unless all ionization states are observed. Photoionization modeling is applied during earlier stages when a central ionizing source remains active and assumes a steady flux of ionizing photons. This technique is used when significant ionization corrections are necessary. 

Several key sources of uncertainty in abundance derivation are discussed in detail by \cite{jose2008_textbook}. One key factor contributing to these errors is the inability of most photoionization models to account for stratified or fragmented ejecta. Previous studies have shown that nova ejecta are often highly fragmented \citep{shore_2016}, resulting in a non-uniform distribution of emission. The filling factor, which accounts for these structural complexities is often treated as a free parameter, with simplified models using integrated line fluxes without incorporating information from line profiles or detailed geometrical configurations. Additionally, despite observational evidence for axial symmetry in some ejecta \citep{Naito_2022}, many photoionization models still assume spherical symmetry. Some systematic uncertainties may contribute to the errors associated with the determined abundances. For example, \cite{andrea1994} used Ionization Correction Factors (ICFs) to convert ionic abundances to elemental abundances. This is commonly done to account for ionic abundances in unobserved ionization stages. Their analysis yielded higher electron densities compared to other studies, which, when combined with the standard ICF method, resulted in larger ionic abundances. For a more in-depth discussion on this one may refer to \cite{schwarz_20002}.

\section{Nova models} \label{sect: models}

For simulations of nova TNRs reaching different peak temperatures of H burning, we use the Nova Framework \citep{pavel2014} to create multi-zone models of Carbon-Oxygen (CO) and Oxygen-Neon (ONe) novae for five different combinations of WD mass, central temperature, and accretion rate (see Table \ref{table:nova mods} for a summary of model parameters and Table \ref{tab:appendB} in Appendix \ref{sect: AppendixB} for a summary of the initial abundances of each model). The Nova Framework \citep{pavel2014} involves using the stellar evolution code \texttt{MESA} \citep{MESA_first,MESA_second} and the multi-zone post-processing nucleosynthesis code of NuGrid \citep{NuGRID,pignatari2016}. The \texttt{MESA} code is used in the Nova Framework to compute the 1D evolution of nova models during their accretion, explosion, and nova-envelope early expansion phases. \null The \texttt{star} module in \texttt{MESA} handles the stellar evolution calculations while supporting modules provide numerical algorithms for adaptive mesh refinement, modern input physics, time-step control, and atmospheric boundary conditions. The input physics for \texttt{MESA} includes tables of opacities, equations of state, and nuclear reaction rates. This work uses the same inputs as those described in \cite{pavel2013,pavel2014}. The \texttt{MESA} output files with temperature, density, radius, and diffusion coefficient profiles as functions of time and mass coordinate are then used in post-processing nucleosynthesis computations done with the NuGrid Multi-zone Post-Processing Nucleosynthesis Parallel code (\texttt{MPPNP}). In \texttt{MPPNP}, the reaction network includes nuclear reaction rates compiled from various sources depending on the mass region. For weak interactions, rates are taken from standard NuGrid libraries, including those by \cite{Fuller1985,oda1994,Langanke2000,Goriely}. For more details see \cite{pavel2014}. The Nova Framework, is now part of the CaNPAN computational tools\footnote{\url{https://github.com/dpa1983/canpan_projects/blob/main/README.md}}. Results of our post-processing nucleosynthesis computations for the multi-zone nova models are compared with observations in Section \ref{sect: comp w obs} to investigate this Ca abundance discrepancy. 

\begin{table*}[htb!]
\centering
\caption{WD type, mass, central temperature, accretion rate, and peak TNR temperature of our multi-zone nova models.}
\label{table:nova mods}
\begin{tabular}{cccccc} 
 \hline
 Nova Model & WD Type & $\text{M}_{\text{WD}}$[\text{M}$_{\sun}$] & $T_{\text{WD}}$[$10^6$K] & $\text{M}_{\text{acc}}$[$\text{M}_{\sun}\,\text{yr}^{-1}$] & $T_{\text{max}}$ [$10^6$K] \\ [0.5ex] 
 \hline
 1 & CO & 1.15 & 12 & $2 \times 10^{-10}$ & 232 \\
 2 & CO & 1.15 & 10 & $10^{-11}$ & 253 \\
 3 & ONe & 1.15 & 12 & $2 \times 10^{-10}$ & 261 \\
 4 & ONe & 1.3 & 20 & $2 \times 10^{-10}$ & 321 \\
 5 & ONe & 1.3 & 7 & $10^{-11}$ & 404 \\ [1ex] 
 \hline
\end{tabular}
\end{table*}

The adopted accretion rates are selected to explore regimes that yield the highest TNR peak temperatures, which are favored in systems with massive WDs, low central temperatures, and slow accretion \citep{truran2008}. Lower accretion rates lead to more massive accreted envelopes and more degenerate ignition conditions, resulting in more energetic outbursts. These rates also align with those used in prior studies on which this work builds \citep{pavel2014}. Among our models, Nova Model 5 achieves the highest peak temperature of $4.04\times10^{8}$ K and is hereafter referred to as the hottest multi-zone nova model.

Mixing plays a key role in nova models in two distinct ways. First, there is mixing within the accreted envelope itself. In \texttt{MESA}, and subsequently in \texttt{MPPNP}, this is treated as time-dependent mixing between mass zones, and is modeled as a diffusive process, where the diffusion coefficient is derived from mixing-length theory. The diffusion coefficient can then be used to account for mixing in the multi-zone post-processing. Second there is mixing between the WD and the accreted material is thought to result from hydrodynamic instabilities \citep{casanova2010,casanova2011}. \null \texttt{MESA} can account for this mixing through convective boundary mixing (CBM), which is treated as a time-dependent diffusive process. In the Nova Framework, CBM is modeled as exponential convective overshooting with the e-folding length scale $f=0.004$ of the pressure scale height \citep{pavel2014}. However, this approach is computationally expensive and instead can be replicated by assuming the accreted material is pre-mixed, with a composition of 50\% solar-like material from the companion star and 50\% material from the WD’s outer layers. This assumption is supported by spectroscopic observations of high metallicities in nova ejecta \citep{gehrz1998}. \cite{pavel2014} demonstrated that multi-zone nova models using this pre-mixed prescription produce similar peak temperatures, rise times and final elemental abundances in nova envelopes similar to those obtained with CBM.

\citet{Kelly2013} investigated one-dimensional hydrodynamic models of ONe novae and identified key elemental abundance ratios that serve as indicators of the degree of mixing. The authors assessed the sensitivity of these indicators to nuclear reaction rate uncertainties using MC methods. By comparing model predictions with observed abundances, they concluded that a 25\% WD to 75\% solar accreted material mixing ratio provides a better match to observations than the commonly adopted 50/50 prescription. We did test a 25/75 WD-to-solar mixing prescription and found that it did not reproduce the low C abundance seen in some observations used in this work nor did it reproduce the high Ca abundance. While the \cite{Kelly2013} result motivates considering alternatives to the 50/50 assumption, our calculations indicate that a lower WD contribution cannot reproduce the unusually high Ca abundances seen in some novae. Given our focus on exploring the most extreme scenarios capable of producing Ca, we adopt the 50/50 pre-mixed prescription. It is beyond the scope of this work to speculate on the reasons behind the potential peculiarities of this observational sample, although we note that the high Ca abundances necessary for detection may be accompanied by atypical abundances of other elements. Further observational investigation is needed. 

While this work focuses on 1D simulations, other studies of novae have employed a 3D approach. For instance, \cite{giovani_2019} conducted an in-depth study of the effects of turbulent mixing on light element synthesis in novae. The author combined results from 3D dynamical simulations, 1D hydrodynamical profiles, and a post-processing approach incorporating a stochastic simulation algorithm to investigate turbulent mixing and nuclear reaction rates. While 3D models simulate mixing more realistically, the validity of 1D approaches has been demonstrated by \cite{pavel2014}, who compared 1D models with the results of a 1D hydrodynamical code, finding good agreement between the two.

More recently, \cite{jose_2020} have combined 1D and 3D modeling techniques to better understand the nova eruption. The authors use a 1D hydrodynamic code to first model the early stages of the explosion such as mass accretion and the beginning of the TNR. Once convection occurs and extends throughout the entire envelope, 3D simulations are used to extract both the amount of mass dredged up from the WD and the convective velocity profile. This information is then fed back into the 1D simulation to complete modeling the explosion through the envelope's expansion and ejection. The researchers compare these combined 1D and 3D mixing results with 1D models computed using pre-mixed accretion material composition, where the amount of WD material is determined from the mean, mass-averaged metallicities in the ejecta obtained from both modeling approaches. Their findings show that more massive envelopes develop in the combined 1D and 3D simulations compared to the purely 1D models with pre-mixed material. 

We acknowledge the existence of nova models developed by researchers at Arizona State University \citep{starrfield2020,starrfield_ONe}, which simulate TNRs on both CO and ONe WDs using the 1D hydrodynamic code \texttt{NOVA}. These models assume that mixing between the accreted and core material occurs after the onset of the TNR, based on results from multi-dimensional studies that show convective instabilities dredge up core material during the explosion phase. These simulations demonstrate enrichment of the ejecta in key radioactive isotopes such as \(^{7}\text{Be}\), \(^{22}\text{Na}\), and \(^{26}\text{Al}\), and argue that novae are important sources of Galactic \(^{7}\text{Li}\) and potentially evolve into Type Ia supernovae (for CO WDs) or neutron stars via accretion-induced collapse (for ONe WDs). However, in the absence of studies directly comparing their nucleosynthetic yields with observed nova abundances, these models are less suitable for our purposes. In contrast, the models developed by \cite{jose1998} include detailed comparisons with observational data and are therefore more directly comparable to our modeling approach. For these reasons, we adopt them as the basis for comparison in this work.

\begin{figure}[htb!]
    \centering
    \includegraphics[width=\columnwidth]{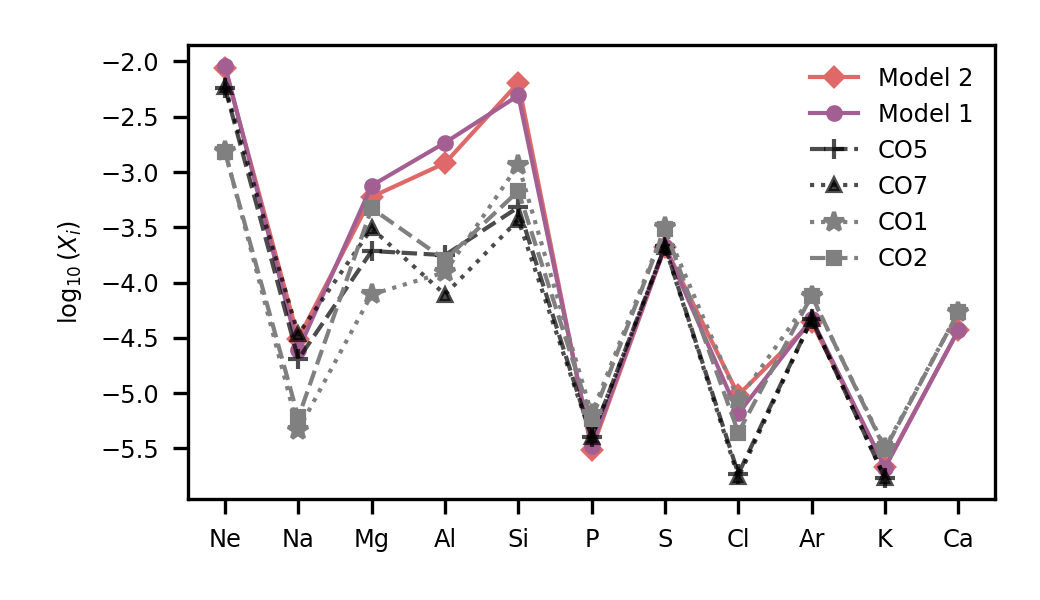}
    \label{fig:ne_bg1}
    
    \includegraphics[width=\columnwidth]{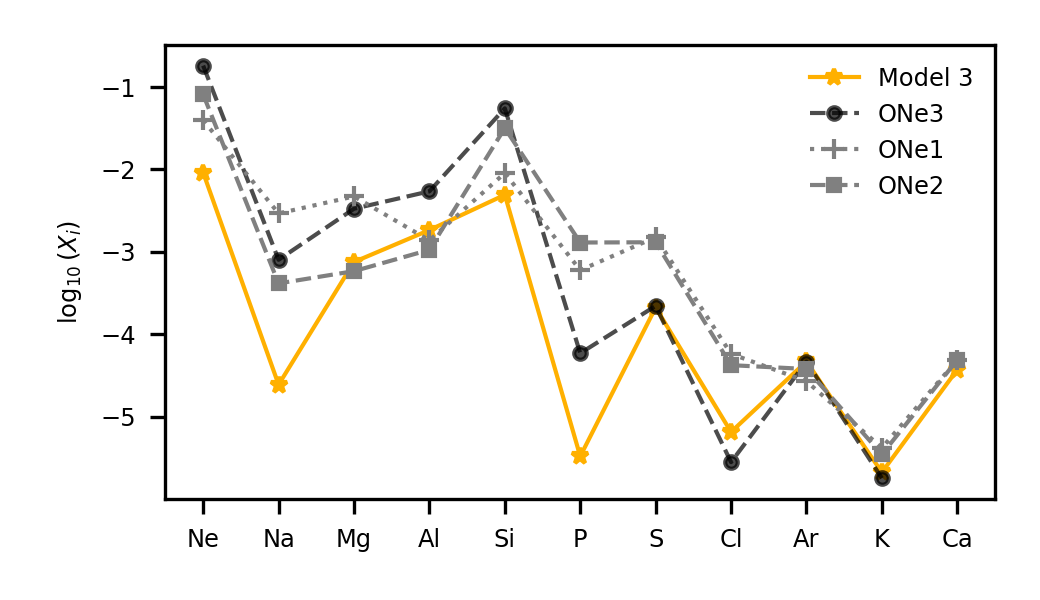}
    \label{fig:ne_bg1_test}
    
    \includegraphics[width=\columnwidth]{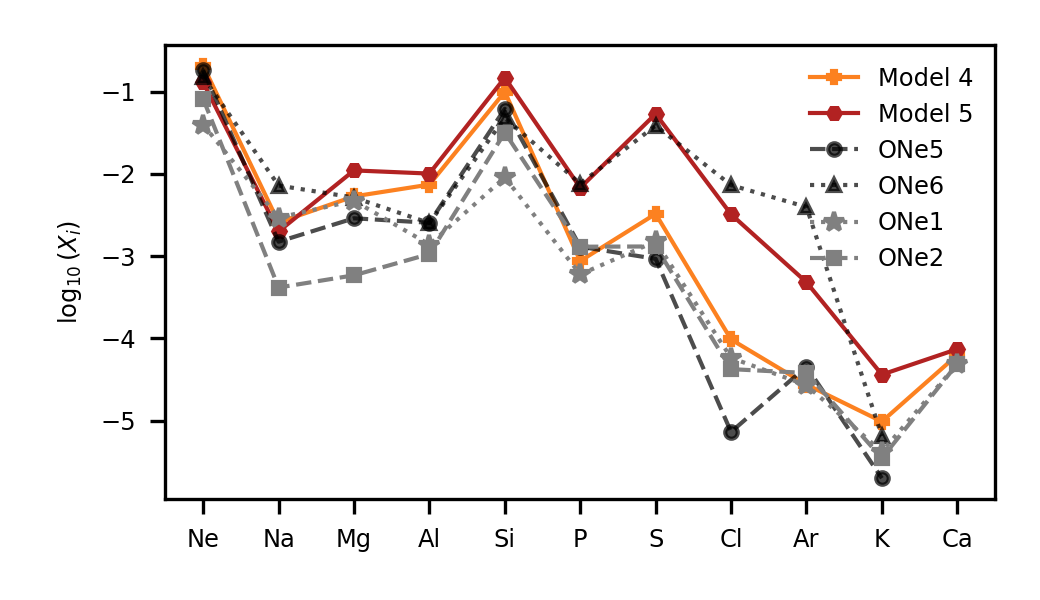}
    \label{fig:ne_bg2}
    
    \caption{Comparison of elemental mass fractions from our multi-zone nova models (\texttt{MESA} and \texttt{MPPNP} simulations in the Nova Framework, solid-color lines) with those of the Barcelona Group (black dashed and dotted lines). Models selected from the Barcelona group have parameters closest to ours for both CO and ONe novae \citep{jose1998}. The gray dashed and dotted lines represent more recent models from \cite{jose_2020}.}
    \label{fig:BG}
\end{figure}

The nova models presented by \cite{jose1998}, hereafter referred to as the Barcelona Group, have parameters similar to ours. Their models show good agreement with observations of elemental abundances in various novae for lighter-mass elements: from H to Ne. In this work, we are interested in the synthesis of elements near and up to Ca; therefore, we compare our results to theirs for elements beyond Ne in \Fig \ref{fig:BG}. The models presented in this work show good agreement with those from the Barcelona group. Any discrepancies are likely to be caused by updated reaction rates in our network, slightly different initial chemical compositions of the accreted mixture \citep[see \Fig 4 of][]{pavel2014}, and differences in the maximum WD masses. For example, model ONe6 of the Barcelona Group has a WD mass of 1.35 M$_{\sun}$, which is closer to the Chandrasekhar mass limit for WDs than the maximum WD mass of 1.3 M$_{\sun}$ of our ONe nova models. We also compare with their more recent models of classical novae from \cite{jose_2020}. Given that the results from the Barcelona Group demonstrated good agreement with observations of lighter mass elements in novae, and the comparison of our nova models with theirs shows good agreement for heavier elements, we can infer that nova models can be reliably compared to observations. Therefore, we can directly compare our models with the observations of Ca abundances in novae.

\section{Comparison with observations} \label{sect: comp w obs}

We have gathered observational data for novae with measured Ca abundances. To see how overabundant the elements are compared to solar we compare abundances from our models and the observations in the standard stellar spectroscopy bracket notation $[X_{\text{i}}/X_{\text{H}}]$. The solar abundances used in the calculation of $[X_{\text{i}}/X_{\text{H}}]$ are taken from \cite{grevesse1993}. Abundances from nova simulations are typically reported as elemental or isotopic mass fractions and plotted relative to solar values \citep{jose1998,jose_2020}, while isotopic abundances are used for investigation of pre-solar grains of purported nova origin \citep{novagrains1}. However, calculating $X_{\text{i}}/X_{\text{i},\sun}$ from ratios of number densities requires us to assume the abundance of H in the nova ejecta, which varies from star to star and is difficult to measure observationally \citep{gehrz1998}.

\begin{figure}[htb!]
    \centering
    \includegraphics[width=\columnwidth]{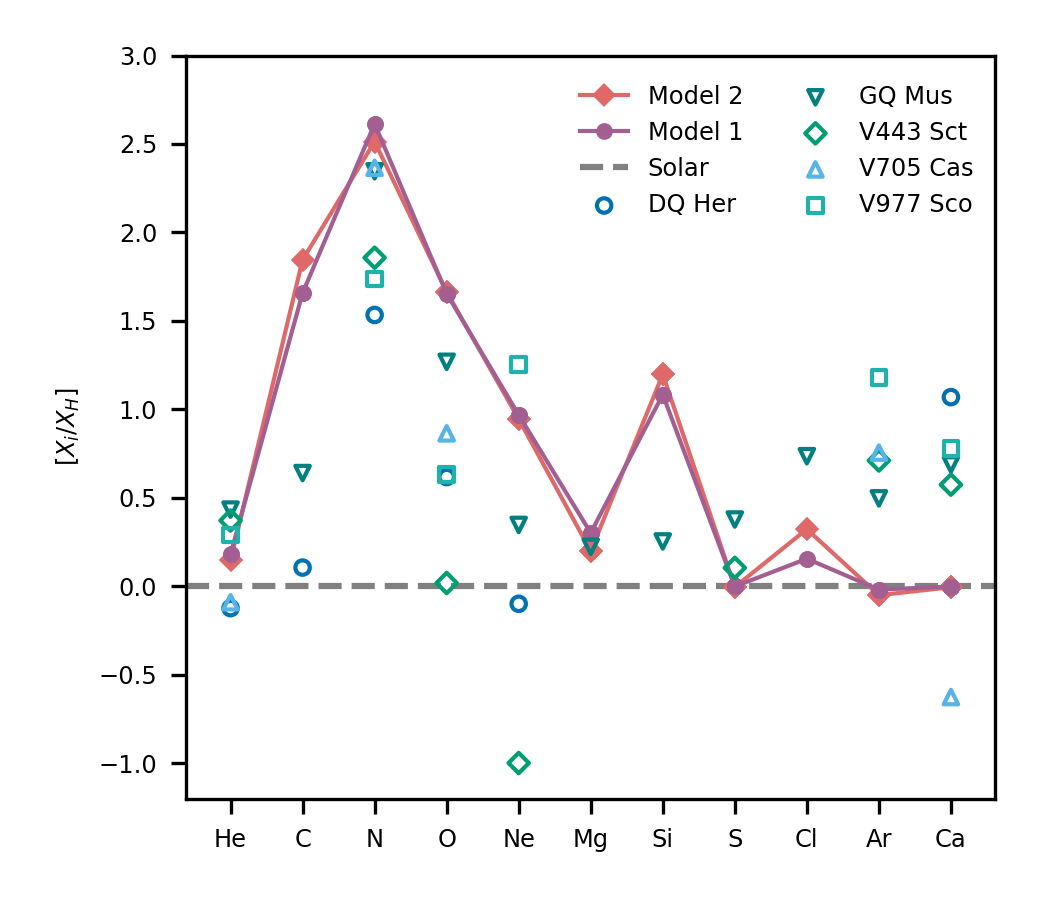}
    \caption{Comparison of observed abundances in CO novae V705 Cas \citep{arkhipova2000}, GQ Mus \citep{morisset1996}, V977 Sco \citep{andrea1994}, V443 Sct \citep{andrea1994}, and DQ Her \citep{Pottasch1959} with predicted abundances from Model 1 and Model 2. Observational data are shown with blue and green symbols, while model predictions are represented by pink and purple circles and diamonds connected by a solid line.}
    \label{fig:co_obs}
\end{figure}

\begin{figure}[htb!]
    \centering
    \includegraphics[width=\columnwidth]{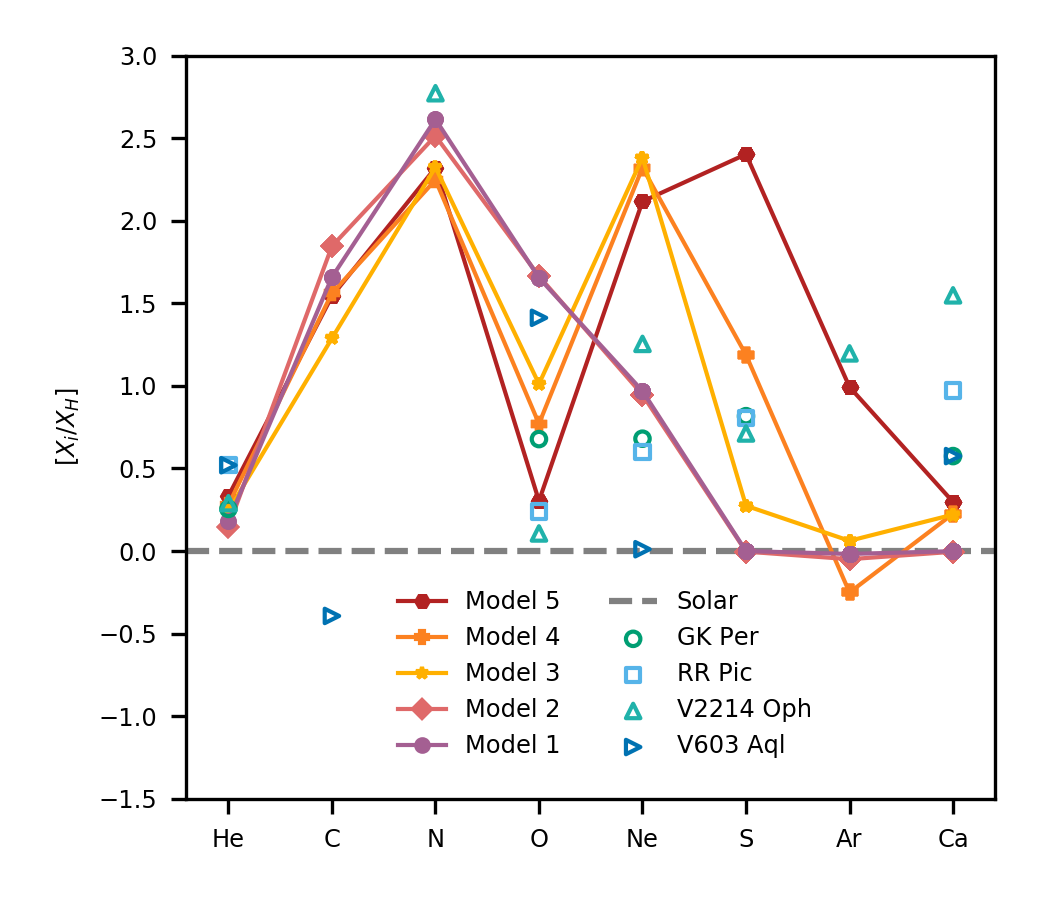}
    \caption{Comparison of observed abundances in novae of uncertain type (CO or ONe) with predicted abundances from multi-zone nova models. Observational data for V1224 Oph \citep{andrea1994}, V603 Aql \citep{Pottasch1959}, RR Pic \citep{Pottasch1959}, and GK Per \citep{Pottasch1959} are shown with blue and green symbols. Model predictions are represented by purple, pink, yellow, orange, and red symbols connected by solid lines.}
    \label{fig:ne_obs}
\end{figure}

To accurately represent nucleosynthesis in our nova models, we must account for H depletion. Our results are presented as [X$_{\text{i}}$/X$_{\text{H}}$], which can inaccurately represent X$_{\text{i}}$. This is because H gets depleted through H burning, meaning that [X$_{\text{i}}$/X$_{\text{H}}$] can appear larger because the denominator has decreased. To address this, we subtract the logarithm of the ratio of H in the pre-mixed material to H in the envelope at the end of the simulations from our results. This adjustment quantifies H depletion and more accurately reflects the elements synthesized during the explosion. In Figures \ref{fig:co_obs} and \ref{fig:ne_obs}, the model data have been downshifted by this factor to highlight changes resulting from nucleosynthesis. For a detailed explanation of this procedure, see Figure \ref{fig:app1} in Appendix \ref{sect:AppendixA}.

In \Fig \ref{fig:co_obs}, the abundances of elements from He to Ca are plotted for Models 1 and 2 and compared against observations of novae that have been classified by observers as the CO type. The mass fraction of H used to calculate [X$_{\text{i}}$/X$_{\text{H}}$] is also taken from the mass averaged surface composition of the model nova envelope. For Model 1, this value is $X_{\text{H}}=0.26$, for Model 2, this value is $X_{\text{H}}=0.27$, with the initial value of H in the 50\% pre-mixed accreted envelope being $0.35$. As shown in \Fig \ref{fig:co_obs}, no Ca is produced in these models, which is expected for CO novae. However, the abundances of Ca from observations of CO novae exceed by nearly one order of magnitude both the solar and our predicted Ca abundances. Considering that CO novae reach lower peak TNR temperatures the presence of considerable overabundances of heavy elements in CO novae is unexpected. Furthermore, Ar also appears to be overabundant in these CO novae, which will be discussed in more detail in Section \ref{sect:disc}.

In \Fig \ref{fig:ne_obs}, all of our multi-zone nova models are compared with observations. In these observations, novae were not explicitly classified by observers as either CO- or ONe-type novae. For the ONe nova models in this figure, the mass fraction of hydrogen used to calculate [X$_{\text{i}}$/X$_{\text{H}}$] was 0.26 for Model 3, 0.23 for Model 4, and 0.16 for Model 5. The observations of Ca seem to be overabundant compared to all models, including our hottest nova model that produces the largest amount of Ca. Due to the very low accretion rate of Model 5 and, as a result, long period between its subsequent explosions, it is highly unlikely that such a nova would be observed. Therefore, it is surprising that the observations show Ca abundances that are much higher than in our hottest nova model. This discrepancy may indicate that there is a significant error associated with the observations of Ca, and probably other heavy-element abundances around it. If the error of these measurements is indeed large, then there could be agreement between the observations and model predictions within those large error bars. Additionally, the majority of the observational data used for abundance comparisons in this study were derived using the nebular analysis method. Specifically, three of the four primary sources in our dataset \citep{andrea1994, arkhipova2000, Pottasch1959} rely on nebular analysis techniques, while only one \citep{morisset1996} employs photoionization modeling. In total, 50 out of 61 individual abundance measurements were obtained via nebular analysis. Since this technique typically yields lower limits on elemental abundances unless all ionization states are accounted for the observed discrepancy between model predictions and observations may in fact be underestimated. Thus, our findings likely represent a conservative estimate of the true discrepancy.

\section{Impact of Nuclear Physics Uncertainties} \label{sect:nphys}

Charged-particle reaction rates and their uncertainties play a crucial role in shaping nucleosynthetic yields for theoretical models of novae. We perform a MC simulation in which selected reaction rates are varied within prescribed uncertainty limits to assess the sensitivity of Ca production to these nuclear inputs.

To model the effect of these uncertainties, each reaction in our nuclear network is assigned a maximum variation factor \( f \), which defines the interval \([1/f, f]\) within which the default rate may be scaled. For each reaction, a scaling factor \( r \) is generated as follows: a random number \( x \) is drawn from a uniform distribution between \( 1 \) and \( f \) in linear space. Then, with 50\% probability, the multiplier is set to \( r = x \), increasing the rate, or to \( r = 1/x \), decreasing the rate. This produces a uniform probability density in linear space across \([1, f]\) for upward variations and \([1/f, 1]\) for downward variations, resulting in equal likelihood of upward or downward changes within the specified bounds. The modified rate is given by \( r \times \text{default rate} \). The reaction rates are varied relative to their NuGrid default values\footnote{The NuGrid default reaction rates for \(^{37}\text{K}(\textit{p},\gamma)^{38}\text{Ca}\) and \(^{38}\text{K}(\textit{p},\gamma)^{39}\text{Ca}\) were replaced with rates from the STARLIB library \citep{starlib} and tested on our hottest nova model. There was a negligible change in the predicted composition. This substitution was necessary to correct a physical inconsistency introduced by a transcription error in the original rates.}, most of which are taken from JINA Reaclib\footnote{\url{https://reaclib.jinaweb.org}}.

The main outcome of the MC simulation is a dataset comprising of distinct mass fraction sets for each element and isotope participating in nova nucleosynthesis. These datasets are analyzed to identify reactions whose rate uncertainties have the strongest impact on the predicted abundances of selected elements or isotopes.
In this analysis, we calculate the Pearson product-moment correlation coefficients, $r_{\text{P}}$, that quantify the strength of the relationship between the changes in reaction rates and variations of abundances they produce, similar to how it was performed in \cite{pavel2021_MC}. However, a strong correlation alone does not guarantee a significant impact on the abundance. To account for this, we introduce a sensitivity parameter, \(\zeta\), defined as the slope of the best-fit line relating the reaction rate variation factor to the resulting abundance relative to the default value for all MC runs. Both quantities are treated linearly, consistent with how the correlation coefficients are calculated. Thus, $\zeta$ directly measures how much the abundance ratio changes per unit change in the rate variation factor. A reaction may be highly correlated with an element’s production, but if $\zeta$ is small, even large changes in the reaction rate will have little effect on the final abundance.

\subsection{Multi-zone Monte Carlo Simulation For The Hottest Nova Model}\label{subsec:multi-zone mc}

For the first time, we performed a reaction rate uncertainty study on a multi-zone nova model by running MC post-processing nucleosynthesis simulations using the \texttt{MPPNP} code \citep{pavel2014, pavel2021_MC}. \texttt{MPPNP} simulations are more physically realistic as these simulations account for time-dependent mixing through detailed radial profiles of temperature, density, and diffusion coefficient, while single-zone simulations only use a temperature and density trajectory and do not consider mixing. One-zone simulations are faster and less computationally expensive and can be used for impact studies to estimate the importance of nuclear physics uncertainties. However, for a more confident analysis of nucleosynthesis in stars, we need to include all the physics relevant to this process, which is why multi-zone simulations are preferred for both individual and MC simulation runs and for comparison with observations. 

We calculate one set of multi-zone MC simulations that varies (\textit{p},$\gamma$), (\textit{p},$\alpha$), ($\alpha$,\textit{p}), and ($\alpha$,$\gamma$) reactions from \(^{2}\text{H}\) to \(^{52}\text{Ti}\) and uses the maximum reaction rate variation factors from the STARLIB library \citep{starlib}. \texttt{MPPNP} is run 1000 times, each with a unique reaction network featuring the randomly varied rates for selected reactions. On eight core processing units, a single-zone calculation takes a few minutes to run, whereas one \texttt{MPPNP} simulation usually takes a few hours. For this reason, we compute 1000 \texttt{MPPNP} simulations. We chose our hottest nova model for these simulations as it produces the most Ca compared to other models.

Table \ref{table:multi-zone mc full} reveals which reaction rate variations, $f_i$ has the strongest impact on the predicted abundance ${X_k}$, relative to their default value, ${X_{k,0}}$. The last two columns in this table contain the corresponding Pearson correlation coefficients, $r_{\text{P}}$, and the parameter $\zeta$, which measures the sensitivity of that reaction in producing the desired element.

\begin{figure}[htb!]
    \centering

    \includegraphics[width=1.1\columnwidth]{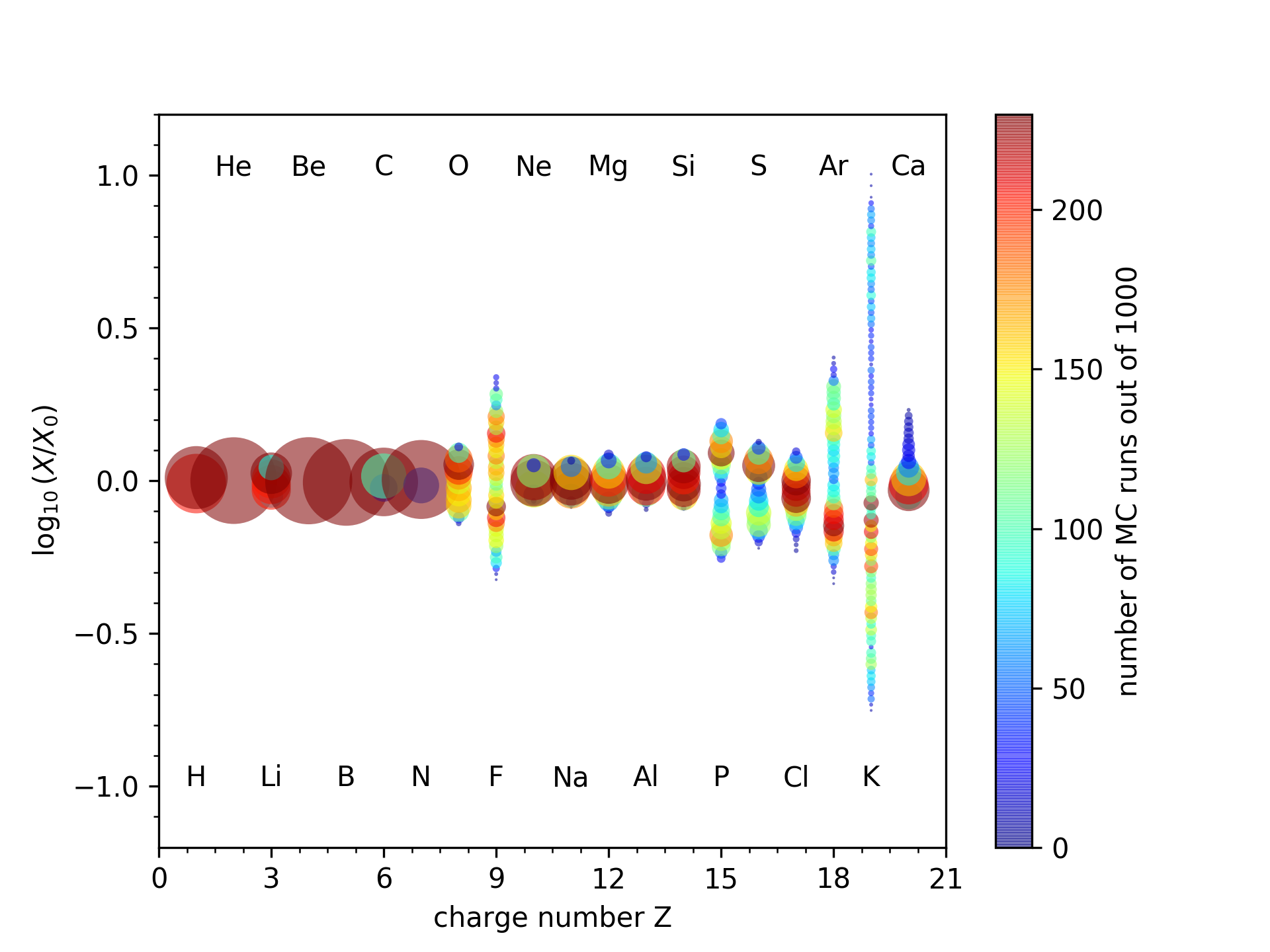}

    \caption{Distributions of the abundances for the selected elements relative to their default value in the multi-zone MC simulation for nova Model 5. The size and color of the circles represent the number of MC runs with that abundance.}
    \label{fig:Multi-zone MC}
\end{figure}

Figure \ref{fig:Multi-zone MC} illustrates the distribution of abundances up to Ca, the color intensity and size of the circles indicate the frequency of abundance occurrences for multiple elements and demonstrates that lighter elements generally exhibit less abundance variation, consistent with well-measured reaction rates for these elements. Conversely, elements near Ca show a larger abundance distribution, reflecting the greater uncertainties in charged-particle reaction rates in this mass region.

Notably, K displays a significant abundance spread in Model 5 when nuclear reaction rates are varied. This could have important implications for the observed K abundance variations in some Globular Clusters (GCs). Recently, several GCs, namely NGC 6715 \citep{Carretta_2022}, NGC 2808 \citep{Mucciarelli_2015}, and $\omega$ Centauri \citep{Garay_2022} have been found to have large star-to-star variations of the K abundance anti-correlating with Mg. This spread in the K abundance as well as the K-Mg anti-correlation could be explained by the same self enrichment process that is believed to be responsible for other proton capture abundance anomalies in GCs (e.g., \cite{pavel_gc}, and references therein). However, as demonstrated by \cite{prantzos_2017}, K can be produced in H burning at the levels reported for the GC NGC 2808 only at temperatures above 180 MK, which are reached during TNRs in novae. \cite{Smith_1996} proposed a scenario in which ONe novae could contribute to the production of the proton-capture abundance variations in GCs. Although ONe novae are currently not considered as the dominant source of these variations, our revealed large uncertainty of the predicted K abundance for ONe novae hints that they still could contribute to the enrichment of stars in some GCs in K.

\begin{table}[htb!]
\centering
\caption{Correlations and sensitivities revealed in the multi-zone Monte Carlo simulation for our hottest nova model. Reaction rates for isotopes from H to Ti were varied with maximum reaction rate variation factors from the STARLIB library \citep{starlib}. Correlations are shown for each element if $\left| r_{\text{P}}(f_i, {X_k}/{X_{k,0}}) \right| \geq 0.15$.}
\label{table:multi-zone mc full}
\begin{tabular}{l c c c} 
\hline
Element & Reaction & r$_{\text{P}}$\tablenotemark{a} & $\zeta$\tablenotemark{b} \\ 
\hline
O & $^{17}$F(\textit{p},$\gamma$) & -0.8150 & -0.7237\\
\multicolumn{4}{c}{}\\[-0.5em]
F & $^{18}$F(\textit{p},$\gamma$) & 0.8772 & 0.8333\\
  & $^{18}$F(\textit{p},$\alpha$) & -0.3186 & -1.076\\
\multicolumn{4}{c}{}\\[-0.5em]
Ne & $^{23}$Mg(\textit{p},$\gamma$) & -0.2992 & -0.1330\\
  & $^{23}$Na(\textit{p},$\gamma$) & -0.2671 & -0.2954\\
\multicolumn{4}{c}{}\\[-0.5em]
Na & $^{23}$Mg(\textit{p},$\gamma$) & -0.3109 & -0.1433\\
  & $^{23}$Na(\textit{p},$\gamma$) & -0.2611 & -0.2820\\
\multicolumn{4}{c}{}\\[-0.5em]
Mg & $^{23}$Na(\textit{p},$\gamma$) & 0.3839 & 0.4904\\
  & $^{26}$Al$^{m}$(\textit{p},$\gamma$)\tablenotemark{*} & -0.2732 & -0.01114\\
\multicolumn{4}{c}{}\\[-0.5em]
Al & $^{23}$Na(\textit{p},$\gamma$) & 0.3221  & 0.3935 \\
  & $^{25}$Al(\textit{p},$\gamma$) & -0.2263  & -0.0682\\
\multicolumn{4}{c}{}\\[-0.5em]
Si & $^{30}$P(\textit{p},$\gamma$) & -0.5124  & -0.02173 \\
  & $^{23}$Mg(\textit{p},$\gamma$) & 0.2207 & 0.1193 \\
\multicolumn{4}{c}{}\\[-0.5em]
P & $^{30}$P(\textit{p},$\gamma$) & 0.8154 & 0.07768\\
  & $^{31}$P(\textit{p},$\gamma$) & -0.2024 & -1.023 \\  
\multicolumn{4}{c}{}\\[-0.5em]
S & $^{30}$P(\textit{p},$\gamma$) & 0.7407 & 0.04905 \\
\multicolumn{4}{c}{}\\[-0.5em]
Cl & $^{37}$Ar(\textit{p},$\gamma$) & -0.4615 & -0.02311\\
  & $^{30}$P(\textit{p},$\gamma$) & 0.3636 & 0.01769  \\
\multicolumn{4}{c}{}\\[-0.5em]
Ar & $^{37}$Ar(\textit{p},$\gamma$) & 0.8863 & 0.1399 \\
\multicolumn{4}{c}{}\\[-0.5em]
K & $^{37}$Ar(\textit{p},$\gamma$) & 0.6219 & 0.3928\\
  & $^{38}$K(\textit{p},$\gamma$) & 0.6005 & 0.3729\\
\multicolumn{4}{c}{}\\[-0.5em]
Ca & $^{37}$Ar(\textit{p},$\gamma$) & 0.4496 & 0.02627 \\
  & $^{38}$K(\textit{p},$\gamma$) & 0.4316 & 0.02479 \\
  & $^{39}$K(\textit{p},$\gamma$) & 0.3419 & 0.1163 \\
\hline
\end{tabular}
\tablenotetext{*}{This is the isomeric state of \(^{26}\text{Al}\).}
\tablenotetext{a}{r$_{\text{P}}$ is the Pearson coefficient estimating the correlation between the reaction rate variation and predicted abundance.}
\tablenotetext{b}{$\zeta$ is a measure of the sensitivity of a given element to its correlated reaction.}
\end{table}

\subsection{Sensitivity of Important Reactions}

Based on the key reactions identified in our multi-zone MC simulation, we investigated their impact on calcium production. \cite{longland1} conducted an in-depth study of the \(^{39}\text{K}(\textit{p},\gamma)^{40}\text{Ca}\) reaction, revealing greater uncertainty in its rate than previously thought, particularly within the temperature range relevant to nova nucleosynthesis.  More recently, \cite{fox2024} proposed that this reaction rate should be increased by a factor of 13 at $7\times10^{7}$ K. In light of these findings, we recomputed our hottest multi-zone nova model, increasing only the \(^{39}\text{K}(\textit{p},\gamma)^{40}\text{Ca}\) reaction rate by a factor of 10 in our nuclear reaction network. The modified simulation resulted in a modest enhancement of Ca production, yet remained insufficient to reproduce the observed abundances. The nuclear physics uncertainties could be a factor of 100, for example, if there is a narrow hidden resonance. For completeness, we also tested this case and found the increase was still minimal compared to the observations. We also tested the \(^{38}\text{K}(\textit{p},\gamma)^{39}\text{Ca}\) reaction, the second most correlated to Ca production, by increasing its rate by factors of 10 and 100 and still saw only a minimal increase in the production of Ca. More importantly, there is a depletion of Ar in both of these scenarios which is an issue since the observations seem to suggest an overabundance of both Ca and Ar.

In Table \ref{table:multi-zone mc full} we see that for Ar, K, and Ca, they all share the same important reactions, namely $^{37}$Ar(\textit{p},$\gamma)$ and $^{38}$K(\textit{p},$\gamma)$. We find that increasing the $^{37}$Ar(\textit{p},$\gamma)$ rate by a factor of 10 and 100 in Model 5 increases the abundance of Ar to better match the observations. This change also slightly increases Ca abundance, but not sufficiently to reach the observed levels. Of the three correlated reactions to the production of Ca, $^{37}$Ar(\textit{p},$\gamma)$ and $^{38}$K(\textit{p},$\gamma)$ have similar correlations and sensitivities (0.02627 and 0.02479 respectively) while \(^{39}\text{K}(\textit{p},\gamma)^{40}\text{Ca}\) has a slightly lower correlation but a much stronger sensitivity (0.1163). Increasing all three rates by factors of 10 and 100 shows that \(^{39}\text{K}(\textit{p},\gamma)\) produces the largest Ca enhancement, consistent with its higher sensitivity, yet still falls short of reproducing the observed abundances.

To test whether nuclear breakout via the \(^{19}\text{F}(\textit{p},\gamma)\) reaction could enhance Ca production in novae, we also increased the branching ratio \(^{19}\text{F}(\textit{p},\gamma)/^{19}\text{F}(\textit{p},\alpha)\) by factors of 10 and 100 in post-processing calculations. Neither modification significantly affected Ca yields, indicating that this reaction pathway cannot explain the observed Ca abundances in novae.

\section{Alternative hypotheses explaining high Ca abundance in novae}\label{sect:disc}

The nuclear physics uncertainties in our nova models, namely the estimated uncertainties of the charged-particle reaction rates, are not be able to account for the discrepancy between observations and our model predictions of the Ca abundance. In this section, alternative explanations of the Ca discrepancy are discussed. 

One possible explanation for the overabundance of Ca in observations is that the accreted material originates from an evolved stellar companion. In this scenario, the accreted nova envelope would be enhanced in heavy elements instead of having solar composition \citep{companion}. It is possible in the Nova Framework to change the initial conditions to reflect such an enriched envelope, which could be the focus of future work. For instance, the effect of increasing \(^4\text{He}\) abundance in the accreted nova envelope has been previously investigated by \cite{pavel2021_Be}. The authors showed that increasing \(^4\text{He}\) in agreement with observational data of \cite{gehrz1998} in nova models could reduce the discrepancy between the observed and predicted abundances of \(^7\text{Be}\) in novae.

\cite{ca_ism} showed that Ca in the interstellar medium (ISM) tends to be converted into dust due to its high condensation temperature compared to other lighter mass elements. If Ca were to become trapped in the dust, which remains after the nova explosion, and these explosions were to happen recurrently, over time it may be possible that Ca builds up around a star relative to other light elements, and thus appears overabundant in observations of novae. However, this theory is inconsistent when considering Ar. With its chemical properties being markedly different from Ca, we would not expect to see an enhancement in Ar in the dust fractionation scenario as it would be blown away with other gases. Therefore, the simultaneous observational overabundance of both Ca and Ar questions dust fractionation as a viable explanation. 

\section{Conclusion}\label{sect: concl}

We have shown that there is a discrepancy between the observed and predicted abundances of Ca and Ar in novae, and have concluded that within the scope of our models, nuclear physics uncertainties cannot account for it. Even the hottest nova model, Model 5, is unable to reach the observed Ca and Ar abundances and because of the very low accretion rate for this model, the probability of observing such novae is low, suggesting that the observations we collected are unlikely to be from novae of this type.

We performed a multi-zone MC simulation for Model 5 to investigate the impact of nuclear physics uncertainties on the charged particle reaction rates. This is the first time such a simulation has been done in the context of nova nucleosynthesis. Using the multi-zone simulations to compare to observations or perform impact studies is crucial because they include more detailed physics than the one-zone simulations, such as mixing, which may play an important role in the nucleosynthesis. 

We found that \(^{37}\text{Ar}(\textit{p},\gamma)\) emerged as the key reaction for both Ca and Ar in this expanded analysis. Subsequently, we increased the reaction rates for \(^{37}\text{Ar}(\textit{p},\gamma)\), \(^{38}\text{K}(\textit{p},\gamma)\), \(^{39}\text{K}(\textit{p},\gamma)\), and \(^{19}\text{F}(\textit{p},\gamma)\) individually by factors of 10 and 100 in our hottest nova model, and none of these rate increases were sufficient to reproduce the  observed abundances in the nova ejecta.

In summary, uncertainties in nuclear reaction rates in proximity to the K-Ar-Ca region of the nuclear chart are unlikely to resolve the discrepancy between observed and predicted Ca abundances in nova ejecta. Further observational and theoretical investigations into the composition of the companion stars, observational effects, and the nova environment are warranted to answer this open question. That being said, experimental measurements of reactions in the region remain important, such as those related to Na or K production, and can help to constrain nova nucleosynthesis models further and relate to observations of pre-solar grains and the study of globular clusters, for instance. It is also important to consider that the observed high Ca abundances may be correlated with other atypical elemental abundances in the observed sites (i.e., low C), leading to the question of whether the observed objects are a self-selecting sample that is not representative of all novae. In concert with advances in modeling, future multi-wavelength observations of a broad sample of nova events will be key to unraveling the mysteries still surrounding these dramatic yet common stellar explosions.

\section{Data Availability}
The inlist files used for our MESA simulations, along with the post-processed data and Jupyter notebooks used for analysis, have been deposited in the MESA Zenodo community at \dataset[doi:10.5281/zenodo.14961565]{https://doi.org/10.5281/zenodo.14961565}.





\section{Acknowledgments}
\begin{acknowledgments}
We would like to extend our warm thanks to the anonymous referee whose comments have significantly improved the quality of this paper. The authors acknowledge generous support from the Natural Sciences and Engineering Research Council of Canada (NSERC) awards: SAPPJ-2023-00039 "Nuclear Astrophysics at TRIUMF-ISAC: Stellar Burning Reaction Studies with DRAGON and TUDA" and SAPPJ-2021-00032  "Nuclear physics of the dynamic origin of the elements". FH acknowledges funding from an NSERC discovery grant. The results of this work were presented at the 2024 Frontiers in Nuclear Astrophysics Conference with funding support from the International Research Network for Nuclear Astrophysics (IReNA). The authors are grateful to Prof. Alan C. Shotter for useful discussions. The numerical simulations for this work were carried out on the Compute Canada Cedar supercomputer operated by WestGrid at the Simon Fraser University and by the Digital Research Alliance of Canada (\url{https://alliancecan.ca/en}). The data analysis was carried out on the Astrohub online virtual research environment (\url{https://astrohub.uvic.ca}), developed and operated by the Computational Stellar Astrophysics group (\url{http://csa.phys.uvic.ca}) at the University of Victoria and hosted on The Alliance Arbutus Cloud at the University of Victoria. 
\end{acknowledgments}

%






\appendix

\section{Adjusting nova models to account for H depletion}
\label{sect:AppendixA}

During our simulations H is depleted via nuclear burning, which can serve as an indicator of the extent of nuclear processing. To accurately represent nucleosynthesis in our nova models, we must account for this hydrogen depletion. Our results are presented as [X$_{\text{i}}$/X$_{\text{H}}$], which can overestimate X$_{\text{i}}$ if H decreases. To address this, we subtract the logarithm of the ratio of H in the pre-mixed material to H in the envelope at the end of the simulations from our results. This adjustment quantifies H depletion and more accurately reflects the elements synthesized during the explosion.

\begin{figure}[htb!]
    \centering
    \includegraphics[width=0.75\textwidth]{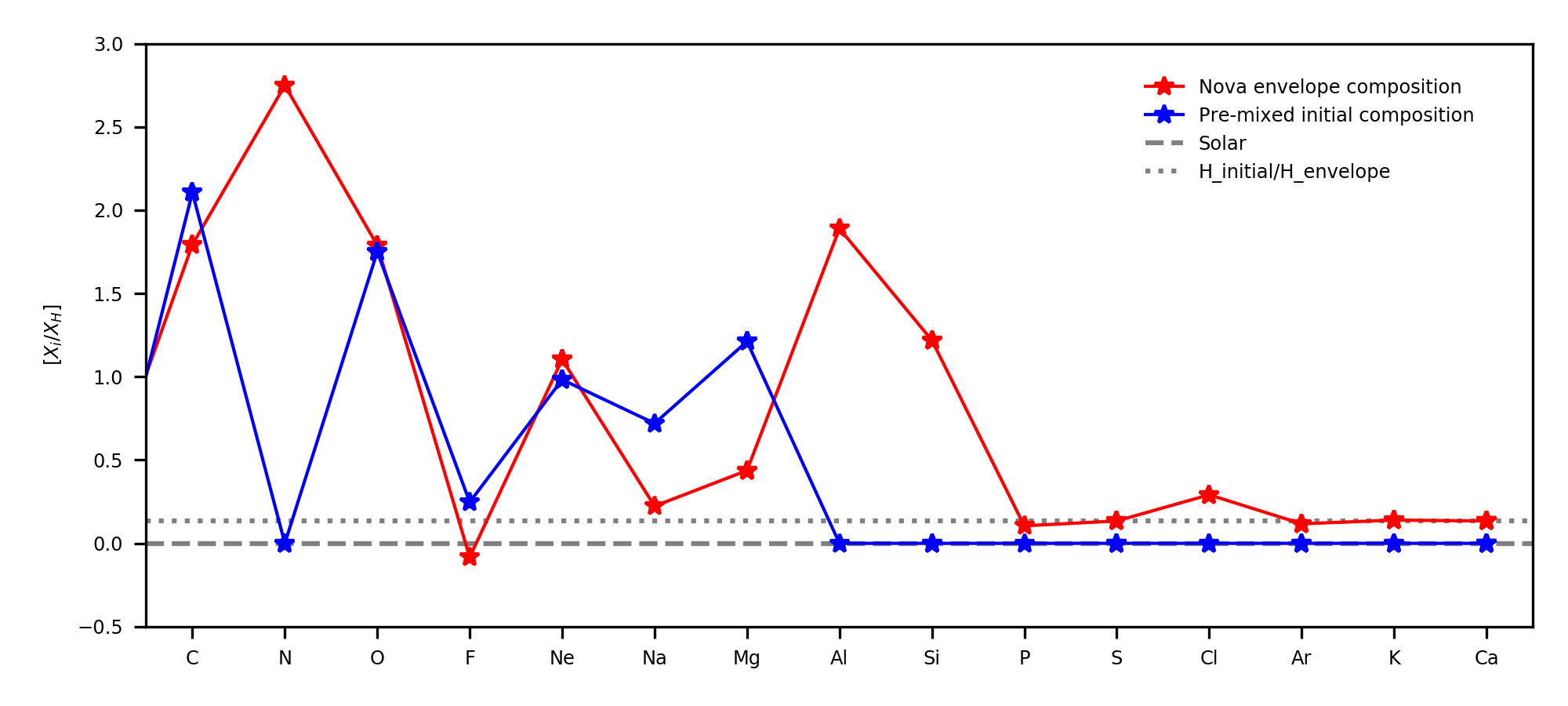}
    
    \includegraphics[width=0.75\textwidth]{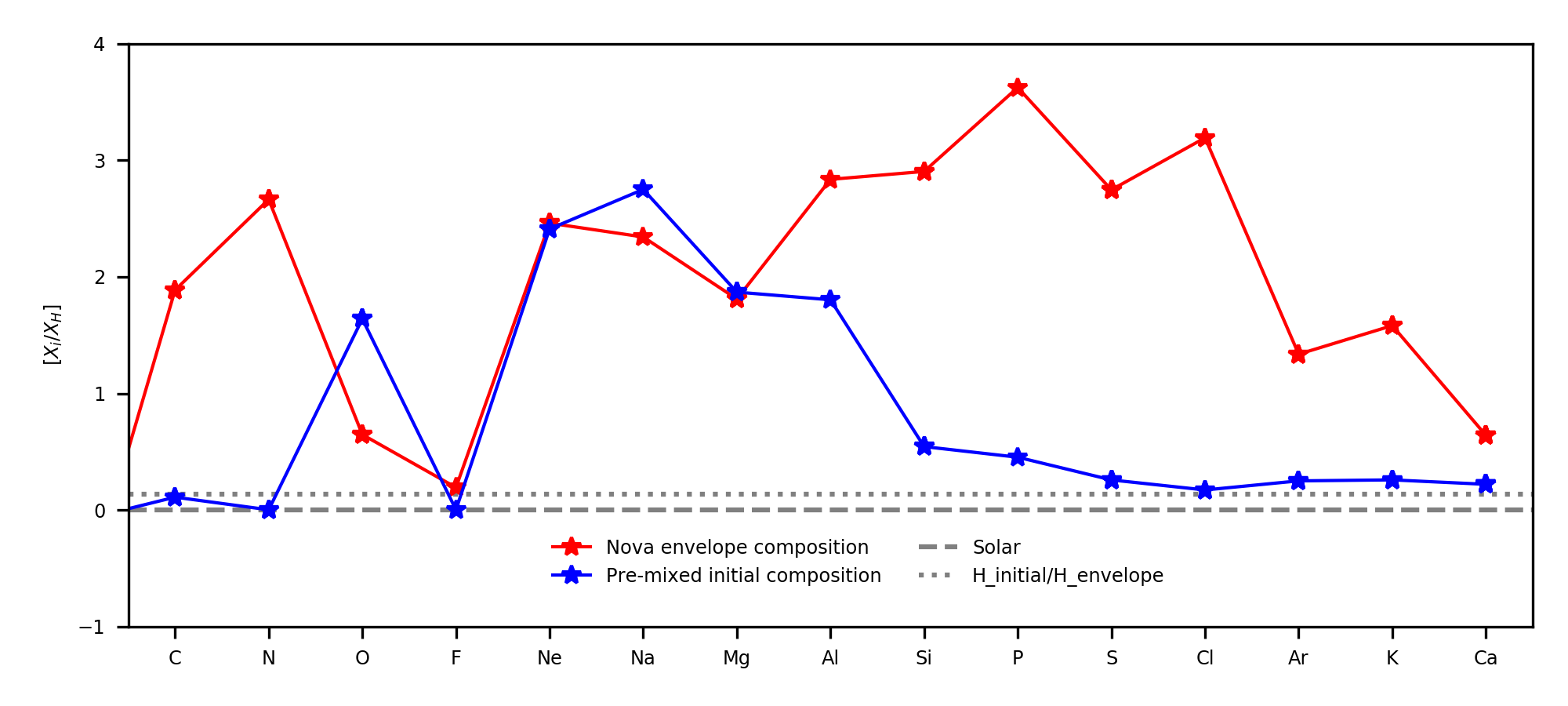}

    \caption{Comparison between the elemental mass fractions of the pre-mixed accreted envelope (blue line) and the post-explosion mass-averaged envelope composition from our multi-zone nova models (red line). The dashed line shows the solar composition and the dotted line shows the fraction of H in the initial composition compared to the H at the end of the simulation. The top panel shows this for Model 1 and the bottom panel shows this for Model 5. For both of the models shown the pre-mixed material is assumed to be 50\% solar material and 50\% WD material. }
    \label{fig:app1}
\end{figure}

Figure \ref{fig:app1} illustrates the distinct impacts of initial mixing and nucleosynthesis on elemental abundances in our analysis. The blue line represents abundances in the pre-mixed material, while the red line shows abundances after the nova explosion. Abundances from the blue line that are above solar levels (dashed line) indicate elements in the pre-mixed material that were already enhanced prior to the nova event. In contrast, abundances from the red line that lie above the dotted line represent elements produced through nucleosynthesis during the nova event. To accurately display which elements are synthesized in the explosion, all models in Figures \ref{fig:co_obs} and \ref{fig:ne_obs} have been downshifted by this H-depletion factor. This approach allows us to clearly distinguish between abundance changes resulting from the use of pre-mixed material and those truly arising from nucleosynthesis during the nova event.

\section{Initial Nova Model Abundances}
\label{sect: AppendixB}

\begin{table*}[htb!]
\centering
\caption{Initial isotopic and elemental abundances from the multi-zone nova models. Elemental abundances are shown in bold and represent the sum of the isotopic abundances. If an isotope is bolded, it indicates that the elemental abundance is given by that single isotope. Models 1 and 2, as well as Models 4 and 5, share the same initial abundances because they have the same WD mass. In contrast, Model 3 is an ONe model with a lower WD mass, resulting in different initial abundances from the other ONe models.} 
\label{tab:appendB}
\begin{tabular}{llcccc}
\hline
& Isotope/Element & Models 1 \& 2 & Model 3 & Models 4 \& 5 & Solar \\
\hline
& $^1$H     & 3.53E-01  & 3.53E-01 &  3.53E-01 & 7.06E-01 \\
& $^2$H     & 6.85E-06  & 6.85E-06 &  6.85E-06 & 1.37E-05 \\
\cline{2-6}
& \textbf{H }       & 3.53E-01  & 3.53E-01 &  3.53E-01 & 7.06E-01 \\
\hline
& $^3$He    & 2.27E-05  & 2.27E-05 &  2.27E-05 & 4.54E-05 \\
& $^4$He    & 1.37E-01  & 1.37E-01 &  1.37E-01 & 2.73E-01 \\
\cline{2-6}
& \textbf{He }      & 1.37E-01  & 1.37E-01 &  1.37E-01 & 2.74E-01 \\
\hline
& $^6$Li    & 2.68E-12  & 2.68E-12 &  2.68E-12 & 5.35E-12 \\
& $^7$Li    & 3.80E-11  & 3.80E-11 &  3.80E-11 & 7.61E-11 \\
\cline{2-6}
& \textbf{Li}      & 4.07E-11  & 4.07E-11 &  4.07E-11 & 8.14E-11 \\
\hline
& \textbf{$^9$Be}    & 5.17E-11  & 5.17E-11 &  5.17E-11 & 1.03E-10 \\
\hline
& $^{10}$B  & 3.21E-10  & 3.21E-10 &  3.21E-10 & 6.42E-10 \\
& $^{11}$B  & 1.43E-09  & 1.43E-09 &  1.43E-09 & 2.86E-09 \\
\cline{2-6}
& \textbf{B}        & 1.75E-09  & 1.75E-09 &  1.75E-09 & 3.50E-09 \\
\hline
& $^{12}$C  & 2.21E-01  & 3.27E-03 &  2.21E-03 & 3.42E-03 \\
& $^{13}$C  & 5.12E-05  & 1.97E-04 &  2.62E-05 & 4.16E-05 \\
\cline{2-6}
& \textbf{C}        & 2.21E-01  & 3.47E-03 &  2.23E-03 & 3.47E-03 \\
\hline
& $^{14}$N  & 5.29E-04  & 5.62E-04 &  5.32E-04 & 1.06E-03 \\
& $^{15}$N  & 2.09E-06  & 2.25E-06 &  2.10E-06 & 4.17E-06 \\
\cline{2-6}
& \textbf{N}       & 5.31E-04  & 5.64E-04 &  5.34E-04 & 1.06E-03 \\
\hline
& $^{16}$O  & 2.71E-01  & 2.14E-01 &  2.10E-01 & 9.62E-03 \\
& $^{17}$O  & 1.91E-06  & 2.37E-06 &  2.03E-06 & 3.81E-06 \\
& $^{18}$O  & 1.09E-05  & 1.43E-05 &  1.09E-05 & 2.17E-05 \\
\cline{2-6}
& \textbf{O}        & 2.71E-01  & 2.14E-01 &  2.10E-01 & 9.65E-03 \\
\hline
& \textbf{$^{19}$F}        & 4.96E-07  & 3.35E-07 &  2.82E-07 & 5.61E-07 \\
\hline
& $^{20}$Ne & 2.63E-03  & 2.34E-01 &  2.49E-01 & 1.82E-03 \\
& $^{21}$Ne & 1.71E-05  & 7.14E-05 &  1.07E-04 & 4.58E-06 \\
& $^{22}$Ne & 6.81E-03  & 5.22E-03 &  4.01E-03 & 1.47E-04 \\
\cline{2-6}
& \textbf{Ne}       & 9.46E-03  & 2.39E-01 &  2.53E-01 & 1.97E-03 \\
\hline
& \textbf{$^{23}$Na}       & 1.05E-04  & 1.78E-02 &  1.12E-02 & 4.00E-05 \\
\hline
&$^{24}$Mg	& 2.28E-03	& 2.68E-02	&  2.46E-02	& 5.86E-04\\
&$^{25}$Mg	& 7.27E-05	& 2.61E-03	&  1.03E-03	& 7.73E-05\\
&$^{26}$Mg	& 3.77E-03	& 2.09E-03	&  2.10E-03	& 8.85E-05\\
\cline{2-6}
&\textbf{Mg}	& 6.12E-03	& 3.15E-02	& 2.78E-02	& 7.52E-04\\
\hline
&\textbf{$^{27}$Al}	& 3.24E-05	& 1.63E-03	& 2.06E-03	& 6.48E-05\\
\hline 
&$^{28}$Si	& 3.73E-04	& 1.07E-03	&  1.36E-03	& 7.45E-04\\
&$^{29}$Si	& 1.96E-05	& 3.13E-05	&  3.15E-05	& 3.92E-05\\
&$^{30}$Si	& 1.34E-05	& 2.53E-05	&  2.48E-05	& 2.67E-05\\
\cline{2-6}
&\textbf{Si}	& 4.06E-04	& 1.13E-03	&  1.42E-03	& 8.11E-04\\
\hline 
&\textbf{$^{31}$P}	& 3.55E-06	& 9.05E-06	&  1.01E-05	& 7.11E-06\\
\hline
&$^{32}$S	& 2.01E-04	& 3.61E-04	&  3.60E-04	& 4.01E-04\\
&$^{33}$S	& 1.63E-06	& 5.90E-06	&  7.01E-06	& 3.26E-06\\
&$^{34}$S	& 9.45E-06	& 1.67E-05	&  1.64E-05	& 1.89E-05\\
&$^{36}$S	& 4.04E-08	& 4.04E-08	&  4.04E-08	& 8.07E-08\\
\cline{2-6}
&\textbf{S}	& 2.12E-04	& 3.84E-04	&  3.83E-04	& 4.23E-04\\
\hline
\end{tabular}
\end{table*}

\begin{table*}[htb!]
\centering
\caption{Initial isotopic and elemental abundances continued.}
\label{tab:appendB2}
\begin{tabular}{llcccc}
\hline
 & Isotope/Element & Models 1 \& 2 & Model 3 & Models 4 \& 5 & Solar \\
\hline
&$^{35}$Cl	& 3.41E-06	& 4.94E-06	&  5.27E-06	& 6.82E-06\\
&$^{37}$Cl	& 1.15E-06	& 1.59E-06	& 1.51E-06	& 2.31E-06\\
\cline{2-6}
&\textbf{Cl}	& 4.56E-06	& 6.53E-06	&  6.78E-06	& 9.13E-06\\
\hline
&$^{36}$Ar	& 4.10E-05	& 7.29E-05	&  7.30E-05	& 8.20E-05\\
&$^{38}$Ar	& 7.87E-06	& 1.41E-05	&  1.40E-05	& 1.57E-05\\
&$^{40}$Ar	& 1.33E-08	& 1.33E-08	&  1.33E-08	& 2.65E-08\\
\cline{2-6}
&\textbf{Ar}	& 4.89E-05	& 8.70E-05	& 8.69E-05	& 9.78E-05 \\
\hline
&$^{39}$K	& 1.95E-06	& 3.53E-06	&  3.66E-06	& 3.90E-06\\
&$^{40}$K	& 2.50E-10	& 2.50E-10	&  2.50E-10	& 5.01E-10\\
&$^{41}$K	& 1.48E-07	& 1.48E-07	&  1.48E-07	& 2.96E-07\\
\cline{2-6}
&\textbf{K}	& 2.10E-06	& 3.67E-06	&  3.80E-06	& 4.20E-06\\
\hline
&$^{40}$Ca	& 3.61E-05	& 6.08E-05	&  6.09E-05	& 7.23E-05\\
&$^{42}$Ca	& 2.53E-07	& 2.53E-07	&  2.53E-07	& 5.06E-07\\
&$^{43}$Ca	& 5.41E-08	& 5.41E-08	&  5.41E-08	& 1.08E-07\\
&$^{44}$Ca	& 8.55E-07	& 8.55E-07	&  8.55E-07	& 1.71E-06\\
&$^{46}$Ca	& 1.71E-09	& 1.71E-09	&  1.71E-09	& 3.43E-09\\
&$^{48}$Ca	& 8.36E-08	& 8.36E-08	&  8.36E-08	& 1.67E-07\\
\cline{2-6}
&\textbf{Ca}	& 3.74E-05	& 6.21E-05	&  6.21E-05	& 7.47E-05\\
\hline
&\textbf{$^{45}$Sc}	& 2.71E-08	& 2.71E-08	&  2.71E-08	& 5.41E-08\\
\hline
&$^{46}$Ti	& 1.62E-07	& 1.62E-07	&  1.62E-07	& 3.23E-07\\
&$^{47}$Ti	& 1.49E-07	& 1.49E-07	&  1.49E-07	& 2.98E-07\\
&$^{48}$Ti	& 1.51E-06	& 1.51E-06	&  1.51E-06	& 3.01E-06\\
&$^{49}$Ti	& 1.13E-07	& 1.13E-07	&  1.13E-07	& 2.26E-07\\
&$^{50}$Ti	& 1.10E-07	& 1.10E-07	&  1.10E-07	& 2.21E-07\\
\cline{2-6}
&\textbf{Ti}	& 2.04E-06	& 2.04E-06	& 2.04E-06	& 4.08E-06\\
\hline
\end{tabular}
\end{table*}

Table \ref{tab:appendB} shows the initial (pre-outburst) abundances for all five nova models. The initial abundance is the composition of the pre-mixed material which is just the sum of 50\% of the WD abundance and 50\% of the solar abundance. The abundances for isotopes are presented followed by the elemental abundance (which is the sum of the isotopic abundances).

\bibliography{ref}{}
\bibliographystyle{aasjournal}




\end{document}